\documentclass[12pt, letterpaper]{article}
\usepackage{amsthm, amsmath, amsfonts}
\def\ds{\displaystyle}
\def\bea{\begin{eqnarray}}
\def\eea{\end{eqnarray}}
\def\be{\bea\begin{array}{c}\ds}
\def\ee{\end{array}\eea}
\begin{document}

\begin{center}

\hfill RUNHETC-07-17\\ 
\hfill ITEP-TH-56/06\\
\vspace{2cm}
\Large{\bf Connecting SLE and minisuperspace Liouville gravity}
\end{center}
\bigskip
\bigskip
\centerline
{\large S. Klevtsov}
\bigskip
\bigskip

\centerline{\it NHETC\,\, and\,\, Department\,\, of\,\, Physics\,\, and\,\, Astronomy,}
\vspace{0.1cm}
\centerline{\it
Rutgers\,\, University,\,\, Piscataway,\,\, NJ\,\, 08854,\,\, USA
}
\vspace{.4cm}
\centerline{\it ITEP\,\, Moscow,\,\,117259,\,\,Russia}
\vspace{.5cm}
\centerline{\small\tt{klevtsov@physics.rutgers.edu}}
\bigskip
\bigskip
\bigskip

\abstract{\footnotesize We show that Fokker-Planck equation for chordal $SLE$ process
under a simple rescaling of the probability density can be traced to
the minisuperspace Wheeler-de Witt equation for boundary operator in 2d Liouville gravity. Insertion of an operator,
calculating $SLE$ critical exponent, corresponds to adding matter
contribution to WdW equation. This observation may be useful for understanding of
why $SLE$ critical exponents are given by KPZ gravitational scaling dimensions. Possible 
applications of the obtained relation are discussed.}

\newpage

\section{Introduction}

Stochastic Loewner evolution ($SLE$), introduced in \cite{schramm}, proved to be
a very useful tool in investigation of random processes on two dimensional plane.
It has been shown that $SLE$ process provides mathematically rigorous framework for 
describing the probability measure on critical curves in lattice models with boundary. In particular, it allows for rigorous
proof of various intersection formulas and determination of 
critical exponents for $c\leq1$ conformal theories \cite{schramm1}.

The chordal $SLE$ procees is defined on the conformal upper-half plane in the following way.
Consider a non-self-intersecting curve
$\gamma_t$ starting at the origin, where $t$ is some parameter along the curve. Let  
$g_t(z)$ be a continuous set of conformal maps transforming
the upper-half plane with a cut along $\gamma_t$ back onto upper-half
plane. Then one can show that it satisfies a simple differential equation
\bea
\label{loewner}
\partial_tg_t(z)=\frac{2}{g_t(z)-a_t},
\eea
first derived by L\"owner in 1923 \cite{loewner}.
To fix the redundant $SL(2,\mathbb R)$ invariance the condition $g_t(z)\approx z+\frac{2t}{z}$ at $z=\infty$ is implied. This choice of boundary condition also encodes a special parameterization of the curve. 
The real function $a_t$ corresponds to image of the end of the curve.
Schramm \cite{schramm} suggested to look at  the equation (\ref{loewner}) as inducing probability measure
on the set of conformal maps $g_t$ and thus on curves $\gamma_t$ from some chosen measure on $a_t$. 
It can be shown that the only choice of the measure for $a_t$, which respects conformal invariance,
reflection symmetry
and produces non-self-intersecting curves, is the one dimensional Brownian motion $\xi_t$  with diffusion
coefficient $\kappa$. 
Defined in this way, 
$SLE$ process provides a natural probability measure on random curves
with many useful properties, see e.g. \cite{gruzberg} for review from a physical perspective.

In the works by Bauer and Bernard \cite{bauer,bauer1} as well as in the subsequent papers \cite{friedrich,cardy,bettelheim}
a nontrivial connection between $SLE$ and conformal field theories with central charge $c\leq 1$ was revealed. In particular, in \cite{bauer} a natural lift of $SLE_\kappa$ to
a formal group formed by exponentiation of lower part of Virasoro algebra
was considered.  It was shown that the martingales (conserved quantities) for this lifted $SLE$ correspond to null vectors in the Verma module of Virasoro algebra, with the CFT central charge related to $\kappa$ by
\bea
\label{central}
c=1-3\frac{(\kappa-4)^2}{2\kappa}.
\eea
However, it seems that the $SLE$/CFT duality cannot explain \cite{bauer} one of the most enigmatic properies of critical conformally-invariant curves, namely the appearance of  KPZ gravitational
scaling dimensions \cite{kpz} as critical scaling exponents. The KPZ formula relates the exponents of random paths in plane geometry
to corresponding
exponents on fluctuating plane. 
This relation has been extensively studied and used to derive exponents of random paths in various lattice models by Duplantier and Kostov \cite{duplantier}, see \cite{duplantier1} for recent review in the context of $SLE$. 
In the framework of $SLE$/CFT duality 
KPZ dimension appears as a scaling exponent of the one point function near the tip of the curve \cite{bauer}. 
It is fair to say, that fundamental reason of why critical exponents of random curves on the plane are given by gravitational dimensions remains to be understood.

One of the goals of this paper is to provide some tentative explanation of this phenomenon in the framework $SLE$. The well known fact, that gravitational dimensions naturally appear 
in the Liouville 2d gravity \cite{kpz}, suggests that there should be some underlying relation between the two subjects. Here observe that $SLE$ process indeed corresponds to the minisuperspace approach to Liouville theory. 
Consider Loewner equation (\ref{loewner}) at some given point on the boundary as an ordinary stochastic Langevin equation.
We show that the corresponding Fokker-Planck equation
after a simple rescaling of probability density takes the form of minisuperspace Wheeler-de
Witt (WdW) equation of 2d gravity \cite{wdw} for the purely gravitational boundary wave function. 
Comparison of two equations leads to following identification of parameters
\bea
\label{main}
\kappa=2\gamma^2,
\eea
where $\gamma$ enters in the Liouville factor of two dimensional metric.
This relation turns out to be in a precise agreement with the relation between $\kappa$ and CFT central
charge (\ref{central}). We also find that mathematical expectation of the operator computing the scaling exponent for
the chordal $SLE$ process \cite{schramm1} satisfies the WdW equation with additional CFT matter. 
The aforementioned rescaling of the probability density has a nice interpretation on the 2d gravity side, where it is nothing but the relation between
vertex operator and wave function. Thus we establish that $SLE$ probability density 
corresponds to the boundary vertex operator. 
Since the minisuperspace approach encodes the KPZ relation in 2d gravity, our observations may shed some light on the appearance of the gravitational dimensions as $SLE$ scaling exponents.

In the minisuperspace approach the equation for gravitational wave function is approximated by only the zero mode part.
Nevertheless this approach is proved to be quite powerful, e.g. it gives correct answer for the exact bulk one-point function
in Liouville theory \cite{fzz}. The WdW zero mode wave function appears also in the matrix model
approach to 2d gravity \cite{wdw}, similar objects has been considered in the $O(n)$ model on random surface
\cite{kostov, kazakov}. The equation we derive here is satisfied rather by
the boundary quantum gravity operator \cite{martinec}, which wave function corresponds to an operator "creating a boundary" in the language of \cite{kazakov}, or the trace of a random path in the language of $SLE$. The bulk gravitational operator then corresponds to an operator "creating a loop" on a random surface. Taking into account the observed relationship between
boundary wave function and $SLE$ trace it might be conjectured, that the bulk wave function corresponds to an $SLE$-type probability measure for random loops.
For $c=0$ such a measure
was constructed by Lawler and Werner (see, e.g., \cite{werner}
and references therein), for generic values of central charge $c\leq1$ the existence of such measure is conjectured in the Malliavin-Kontsevich-Suhov theory \cite{kontsevich}.

This paper is organized as follows. In sec. 2 we review the euclidean quantum mechanics language for
the It\^o calculus and apply it to derivation of Fokker-Plank equation for $SLE$. In sec. 3 we show how WdW equation appears for pure 
$SLE$ process. In sec. 4 we review the calculation
of one-sided scaling exponent and show that it corresponds to WdW
equation with addition of matter. Discussion and comments are presented in the last section.

\section{Quantum mechanics and It\^o calculus}

We start with a brief reminder on the relation between It\^o formulas for
stochastic processes \cite{oksendal} and euclidean quantum mechanics \cite{zinn-justin}.

To define a measure for the stochastic process $y_t$ one connects it to the
Brownian motion $\xi_t$ via the Langevin equation
\bea
\label{lang}
dy_t=u(y_t,t)dt+v(y_t,t)d\xi_t,
\eea
where $u$ and $v$ are some given functions. The probability distribution for Brownian motion can be represented as
path integral in euclidean quantum mechanics
\bea
\label{path}
P_{\xi}(q',0; q,t)=\langle q|e^{-\frac t\kappa
\hat H}|q'\rangle=\int_{(q',0)}^{(q,t)}e^{-\frac1{2\kappa}\int_0^t{\dot\xi_t^2}dt}\mathcal{D}\xi_t.
\eea
Therefore we can use the equation (\ref{lang}) to find the induced
measure on $y_t$ from the brownian measure on $\xi_t$.
To illustrate this consider the example of $SLE$, i.e. set $u=2/y$ and
$v=-1$
\bea
\label{langevin}
dy_t=\frac2{y_t}dt-d\xi_t.
\eea
This is nothing but the $SLE$ equation (\ref{loewner}) written in Langevin form
for the shifted variable $y_t(z)=g_t(z)-\xi_t$ and for some point $z$ on the boundary so that $y_t(z)$ is a real variable.
Now we make the change of variables $\xi\rightarrow y_t(\xi)$ in the path integral (\ref{path})
\bea
\label{path1}
P_y(q',0; q,t)=\int_{(q',0)}^{(q,t)}e^{-\frac1{2\kappa}\int_0^t{(\dot
y-\frac2{y})^2}dt}J[y]\mathcal{D}y,
\eea
where we omitted the subscript $t$ to simplify the notations. Jacobean $J[y]$ comes from the transformation of path integral measure. It can
be computed directly \cite{zinn-justin} and has the form
\bea
\label{jacobian}
J[y]=e^{\int_0^t\frac1{y^2}dt}
\eea
One can also read off its form from the following considerations.
The path integral (\ref{path1}) satisfies the Shr\"odinger equation
\bea
\label{shrodinger}
\frac{\partial}{\partial t}P(y,t)=-\frac12\frac{\partial}{\partial y}\left(\kappa\frac{\partial}{\partial
y}P(y,t)-\frac4yP(y,t)\right),
\eea
where the total derivative on the rhs leads to conservation of the overall probability.
Therefore the role of the Jacobean (\ref{jacobian}) is to adjust the measure in the appropriate way.

The Shr\"odinger equation (\ref{shrodinger}) can be interpreted as the Fokker-Planck equation (backward Kolmogorov equation) in stochastic calculus. 
To treat the expectation values of operators
\bea
\label{mean}
\langle\mathcal{O}(y,t)\rangle=\int dq \langle q|\hat O(\hat q,t)e^{-\frac t\kappa
\hat H}|y\rangle=\int dq O(q,t)\langle q|e^{-\frac t\kappa
\hat H}|y\rangle
\eea
one makes use of Heisenberg representation.
The Heisenberg equation on (\ref{mean}) differs from (\ref{shrodinger}) due to presence of linear
momentum term in the quantum hamiltonian
\bea
\hat H=\frac12\hat p^2+\hat p \frac1{\hat y}+ \frac1{\hat y}\hat p+\frac{\kappa}{\hat
y^2},
\eea
where
\bea
\label{commute}
[\hat p, \hat y]=\kappa,\,\,\,\,\,\,\,\,\hat p= \kappa \frac{\partial}{\partial
y},
\eea
and has the form
\bea
\label{heis}
\left[\frac{\partial}{\partial t}+\frac{\kappa}2
\frac{\partial^2}{\partial y^2}+\frac2y\frac{\partial}{\partial
y}\right]\langle\mathcal{O}(y,t)\rangle=0.
\eea

Stochastic counterpart of the Heisenberg equation is the It\^o formula \cite{oksendal}.
Given a random process (\ref{lang}) any function $h(y_t,t)$ of
random variable $y_t$ and time $t$ satisfies the It\^o equation
\bea
\label{ito}
dh(y_t,t)=dt\frac{\partial }{\partial t}h(y_t,t)+dy_t\frac{\partial }{\partial
y}h(y_t,t)+(dy_t)^2\frac12\frac{\partial^2 }{\partial y^2}h(y_t,t),
\eea
where we assume the following formal multiplication rules for differentials
\bea
\label{ito1}
dt\cdot dt=dt\cdot d\xi_t=d\xi_t\cdot dt=0,\,\,\,\,\,\,\,\,d\xi_t\cdot
d\xi_t=\kappa dt
\eea
Using (\ref{langevin}, \ref{ito}, \ref{ito1}) one can easily check that It\^o formula is
equivalent to (\ref{heis}) after taking the expectation values of operators.

\section{Minisuperspace Wheeler-de Witt equation and $SLE_\kappa$}

Consider the following ansatz for the probability density
\bea
\label{fp1}
P(y,t)=e^{-\frac E\kappa t}P(y)
=e^{-\frac E\kappa t}y^{\frac12+\frac2\kappa}\Psi(y).
\eea
Here $\Psi(y)$ is the (rescaled) stationary wave function in the euclidean quantum mechanics.
Equation (\ref{shrodinger}) now reads
\bea
\label{fp2}
\left[-\left(y\frac{\partial}{\partial y}\right)^2+\frac{2E}{\kappa^2}y^2+\left(\frac12-\frac2{\kappa}\right)^2\right]
\Psi(y)=0.
\eea
In this equation one can recognize the WdW equation in Liouville theory.
Let us briefly remind how it appears in the minisuperspace approach to 2d gravity \cite{wdw}.

Consider the system of Liouville ($\phi$) and conformal matter ($\varphi_i$) fields
with the action
\bea
\label{liouville}
S(\phi, \varphi_i, \hat g)=S_{\mathrm liouv}(\phi, \hat g)+S_{\mathrm matter}(\varphi_i, \hat
g)=\nonumber\\=\frac1{8\pi}\int d^2z\sqrt{\hat g}\left((\hat\nabla\phi)^2+
\frac\mu{\gamma^2}e^{\gamma\phi}+\left(\gamma+\frac2\gamma\right)\phi R(\hat g)\right)+S_{\mathrm matter}(\varphi_i,\hat g),
\eea
where the Liouville field $\phi$ is defined as a scale factor of the
two-dimensional metric $g=e^{\gamma\phi}\hat g$ relative to some reference metric $\hat g$, $\mu$ is the bulk cosmological constant and the background charge $Q=\gamma+2/\gamma$.
Liouville part of the action is a conformal field
theory with central charge $26-c$ if the parameter $\gamma$
is related to the matter central charge $c$ as
\bea
\gamma=\frac1{\sqrt{12}}(\sqrt{25-c}-\sqrt{1-c}).
\eea
The wave functions of the coupled system
factorize onto matter and gravitational part
$\Psi=\Psi^{\mathrm matter}\otimes\Psi^{\mathrm liouv}$. 
Any spinless primary operator of conformal dimension $\delta$ in the bulk (or in the boundary)
can be dressed by the exponent of Liouville field, such that the resulting operator
can be integrated over bulk (resp. boundary)
\bea
\int d^2z e^{\alpha\phi}\Phi_0(z,\bar z),\,\,\,\,\,\,\,\,\,\,\,
\oint e^{\alpha\phi}\Phi_0(x),
\eea
where $\alpha$ is determined by the condition that the sum of CFT and Liouville conformal dimensions equals one
\bea
\label{alpha}
\delta-\frac12\alpha(\alpha-Q)=1.
\eea
The corresponding wave function should therefore satisfy the operator equation 
\bea
\label{wdw}
(L_0^{\mathrm liouv}+\delta-1)\Psi=0.
\eea 
If matter boundary conditions are diffeomorphism invariant, then $\Psi$ should depend on a diffeomorphism-invariant variable. The natural variable to use is the length of the boundary $l=\oint e^{\frac{\gamma}{2}\phi}$. Then in the minisuperspace approximation we restrict to the field configurations, independent of $\sigma$ coordinate on the worldsheet, e.g. to the zero mode part $\phi_0$ of Liouville field. In this approximation (\ref{wdw}) reduces just to $T_{00}\Psi=0$, which is nothing but the Wheeler-de Witt equation, stating that gravitational wave function is invariant under time diffeomorphisms. In this form it is also equivalent to Shr\"odinger equation in Liouville quantum mechanics and $\phi_0$ - to the quantum mechanics coordinate. In terms of the parameter $l=e^{\frac{\gamma}{2}\phi_0}$ the Wheeler-de Witt equation becomes
\bea
\label{wdw1}
\left[-\left(l\frac{\partial}{\partial l}\right)^2+\frac \mu{\gamma^4}l^2+\nu^2\right]\Psi(l)=0,
\eea
where for the boundary operator
\bea
\label{wdw2}
\nu^2=\nu_{boundary}^2=\frac{2\delta}{\gamma^2}+\left(\frac12-\frac1{\gamma^2}\right)^2.
\eea
Originally this equation was written for bulk operators \cite{wdw} 
for which the index $\nu$ doubles: $\nu_{bulk}=2\nu_{boundary}$. Although the boundary WdW equation has not been considered in the literature,
there is some indirect indication that it should have twice smaller index 
compared to bulk equation, i.e. as in (\ref{wdw2}). 

Comparing (\ref{wdw1}) to (\ref{fp2}) we can identify $SLE$ and WdW parameters. First, the length of boundary $l$ corresponds to stochastic variable $y$. This is not surprising since one can look at the $SLE$ process, as inducing random measure on the conformal factor of the upper-half-plane metric $ds^2=|\partial_zg_t(z)|^2\,dzd\bar z$. Second, the "energy" $E$ has to be identified with the cosmological
constant $\mu$. This is also rather natural identification since both parameters can be rescaled arbitrarily: in $SLE$ due to the scaling invariance $y\rightarrow \lambda y$, $t\rightarrow \lambda^2 t$, and in Liouville theory by shifting the field $\phi$ by a constant. Third, the conformal dimension of matter is set to zero
$\delta=0$, i.e we consider pure gravitational wavefunctions in the
Liouville-matter system. Finally, we get
\bea
\label{id1}
\kappa=2\gamma^2.
\eea
This last formula is in precise agreement with the
relation between the CFT central charge $c$ and $SLE$ parameter $\kappa$ (\ref{central}). 
Under these identifications Fokker-Plank equation (\ref{fp2}) takes exactly the form (\ref{wdw1}).
There is also a nice interpretation of the rescaling of probability density (\ref{fp1})
by $y^{\frac12+\frac2\kappa}$. This is equivalent to the relation between boundary Liouville vertex operator and WdW wave function
\bea
V(\phi)=e^{\frac{Q}4\phi}\Psi(l).
\eea
Therefore the $SLE$ probability measure defined by the Fokker-Planck equation (\ref{fp2}) corresponds to purely gravitational boundary vertex operator.

In the next section we extend this correspondence to the operators with nonzero matter conformal dimension $\delta\neq 0$.

\section{$SLE_{\kappa}$ scaling dimensions and WdW}

Various critical exponents associated with random paths on two dimensional plane
have been studied for a long time \cite{duplantier}, see also review \cite{duplantier1} for complete list of references. 
In the context of $SLE$ the critical exponents (crossing probabilities) have been derived in \cite{schramm1}. 

The simplest critical $SLE$ exponent is the so-called one-sided chordal intersection exponent.
We refer to the first paper in \cite{schramm1} for rigorous definition. 
For our purposes it suffices to say that one has to compute the scaling of
expectation value of the derivative 
of the conformal transformation
$y_t'(z)^\delta$ at the point $z$ on the boundary and close to the origin. 
Using (\ref{loewner}) one can easily show that
\bea
y_t'(z)^\delta=\exp\left(-\delta\int_0^t\frac2{y_s^2}ds\right).
\eea
The scaling law of this operator for small $z$ determines the critical exponent $\Delta$
\bea
\label{operator1}
h(y,t)=\mathbf{E}\left[\exp\left(-\delta\int_0^t\frac2{y_s^2}ds\right)\right]\sim \left(\frac y{\sqrt t}\right)^{\Delta(\delta,\kappa)}.
\eea
which is found to be
\bea
\label{index}
\Delta(\delta,\kappa)=\frac{\kappa-4+\sqrt{(\kappa-4)^2+16\delta\kappa}}{2\kappa}=
\frac12-\frac1{\gamma^2}+\sqrt{\left(\frac12-\frac1{\gamma^2}\right)^2+\frac{2\delta}{\gamma^2}}.
\eea
In this expression one can recognize the KPZ gravitational scaling dimension \cite{kpz}. The latter appears in the one-point function of the dressed operator $\Phi_0$ of bare dimension
$\delta$ at fixed length $l$ in Liouville theory 
\bea
\frac{1}{Z(A)}\int\mathcal{D}\phi\mathcal{D}\varphi_is^{-S(\phi, \varphi_i, \hat g)}
\delta\left(\oint ds\, e^{\frac\gamma 2\phi}-l\right)\oint s e^{\alpha\phi}\Phi_0\sim l^{1-\Delta},
\eea
and therefore related to Liouville dressing exponent $\alpha$ (\ref{alpha}) as $\alpha=1-\Delta/\gamma$.

It is easy to derive the  It\^o equation (\ref{ito}) for the expectation value (\ref{operator1}). 
Similar to (\ref{ito}) we arrive at
\bea
\label{pde}
\left[\frac{\partial}{\partial t}+\frac{\kappa}{2}\frac{\partial^2}{\partial
y^2}+\frac{2}{y}\frac{\partial}{\partial y}-\frac{2\delta}{y^2}\right]h(y,t)=0.
\eea
Following the discussion in sec. 2 one can write down the corresponding Fokker-Planck equation,
which is just the Shr\"odinger picture (\ref{mean}) of the previous equation.
Using then the same ansatz as for the wave function in (\ref{fp1}) the equation becomes
\bea
\label{shrodinger1}
\left[-\left(y\frac{\partial}{\partial y}\right)^2+\frac{2E}{\kappa^2}y^2+
\frac{4\delta}{\kappa}+\left(\frac12-\frac2{\kappa}\right)^2\right]
\Psi(y)=0,
\eea
The same identification of parameters as in the previous section
leads us to the minisuperspace WdW equation for the wave function
of the gravitationally dressed boundary operator with a nonzero matter conformal dimension $\delta$
(\ref{wdw1}).

\section{Discussion}

In this paper we considered chordal $SLE_\kappa$ process and showed 
that the Fokker-Plank equation for this process under a simple
rescaling of the probability density takes the form of minisuperspace Wheeler-de
Witt equation for 2d gravity. We also show that insertion of an operator calculating
scaling exponent modifies FP equation in a very same manner as adding
matter contribution to WdW equation. Since the latter encodes KPZ relation this result provides some insight to the appearance of
gravitational scaling dimensions in SLE. Let us now comment on other possible applications of this result. 

Despite being only an approximation, the minisuperspace approach nevertheless gives  
exact answers for some 2d gravity correlators,
e.g. for the one-point function of the bulk Liouville exponent \cite{fzz} in the presence of 
boundary cosmological constant $\mu_b$
\bea
\langle e^{\alpha \phi} \rangle =\frac{U(\alpha |\mu_b)}{|z-\bar z|^{2\Delta_\alpha}} =
\frac1{|z-\bar z|^{2\Delta_\alpha}}\int_0^\infty \frac{dl}{l}e^{-\mu_bl}\Psi_{\alpha}(l).
\eea
Here  $\Psi_{\alpha}(l)$ satisfies eq. (\ref{wdw1}) with $\nu=\nu_{bulk}$ and with the Liouville exponent $\alpha$ related to $\delta$ by eq. (\ref{alpha}).
Equation (\ref{wdw1}) appears also in the framework of random matrix formulation of 2d quantum gravity. In the Hermitean one-matrix model WdW wave function of $(2,2m-1)$ minimal model coupled to gravity arises \cite{wdw} at the $m^{th}$ multicritical point in the double-scaling limit of the "macroscopic loop" operator, creating a hole of boundary length $l$
\bea
w(l)=\frac1{l}\,\mbox{tr}\,\,e^{lM}.
\eea  
The correlation functions of multiple insertions of macroscopic operators have been computed in the matrix model approach \cite{banks,wdw} and are in agreement \cite{martinec1} with exact results, provided by Liouville theory \cite{fzz}. Recently the minisuperspace wave function was also interpreted as ZZ brane amplitude \cite{kutasov}.

In the loop gas approach \cite{kostov,kazakov} similar object appears as the solution to loop equations and has the same interpretation, as an operator creating closed boundary of lenght $l$ on the worldsheet. Boundary operator, creating $SLE$ trace, considered 
in the present paper, seemingly corresponds to the operator creating piece of open boundary at some point of the closed boundary \cite{kazakov}. The loop gas approach
is also in agreement \cite{ponsot} with the boundary two-point function calculation of Fateev, Zamolodchikov and Zamolodchikov and Teschner, thus providing another indication of the validity of minisuperspace approach beyond the semiclassical limit.

Random matrix model interpretation of the boundary operator, as creating
 $SLE$ curve, could provide some insight for the generalization of $SLE$-type measure to random loops. This problem was recently addressed by Lawler and Werner \cite{werner}, and by Kontsevich and Suhov who conjectured the existence of such measure for $c\leq1$ \cite{kontsevich}.
In the light of the relation observed here, one could conjecture that the insertion of bulk operator creating a loop should correspond to $SLE$-type process for random closed curves. Thus the WdW wave function, satisfying eq. (\ref{wdw1}) with index $\nu=\nu_{bulk}$, may have some relevance to (some approximation of) the $SLE$-type probability measure on random loops. 

We obtained the correspondence between $SLE$ and minisuperspace Liouville theory
for the simplest possible $SLE$ scaling dimension. It would be very interesting to show whether there is a deeper connection between the two subjects. In view of this let us mention, that there exists a variety of different critical exponents for $SLE$ process \cite{schramm1}. It would be also interesting to extend this relation to the generalizations of $SLE$, such as $SLE(\kappa,\rho)$ \cite{cardy}. 
Let us also note, that the connection between $SLE$ and 2d Liouville gravity has been discussed from different perspectives in \cite{rbauer,rushkin}.

\vspace{.5 cm}

I am grateful to D.~Friedan for interesting and stimulating discussions. 
I also would like thank A.~Alexandrov, J.~Cardy, M.~Douglas, A.~Gorsky, M.~Kontsevich, D.~Kutasov, A.~Marshakov, A.~Mironov, A.~Morozov,
D.~Shih, Yu.~Suhov, P.~Wiegmann and A.~B.~Zamolodchikov for interesting discussions and comments. Hospitality of I.H.E.S., where final version of this paper was written, 
is gratefully acknowledged.
This research was supported in part by DOE grant DE-FG02-96ER40959, by the grant for support of scientific schools NSh-8004.2006.2, by RFBR grant 07-02-00878 and by the Federal Agency of Atomic Energy of Russia.

\end{document}